\documentclass[twocolumn,showpacs,pra,aps]{revtex4}
\usepackage{dcolumn}
\usepackage{amsmath}
\usepackage{amssymb}
\usepackage{graphicx}
\DeclareMathOperator{\arth}{arth}

\newcommand{\X}{{\mathcal X}}
\newcommand{\T}{{\mathcal T}}
\newcommand{\RR}{{\mathbb R}}
\newcommand{\NN}{{\mathbb N}}
\newcommand{\abs}[1]{\lvert#1\rvert}
\begin{document}

\title{Semiclassical Solution of the Quantum Hydrodynamic Equation for
Trapped Bose-condensed Gas in the $\lowercase{l}=0$ Case} 

\author{Tam\'as Tasn\'adi}

\affiliation{Research Group for Statistical Physics of the Hungarian
  Academy of Sciences,\\ P\'azm\'any P\'eter s\'et\'any 1/A, H-1117
  Budapest, Hungary}

\date{\today}

\begin{abstract}
In this paper the quantum hydrodynamic equation describing the collective,
low energy excitations of a dilute atomic Bose gas in a given trapping
potential is investigated with the JWKB semiclassical method. In the case of
spherically symmetric harmonic confining potential a good agreement is shown
between the semiclassical and the exact energy eigenvalues as well as wave
functions. It is also demonstrated that for larger quantum numbers the
calculation of the semiclassical wave function is numerically more stable than
the exact polynomial with large alternating coefficients. 
\end{abstract}


\maketitle

\section{Introduction}
The main aim of this work is to study the applicability of the JWKB
semiclassical quantization method for the solution of the quantum
hydrodynamical equation describing the collective excitations of a trapped
Bose-condensed gas. Although the analogous equation for homogeneous systems
has been known for long, the equation for trapped gases has first been written
down and solved for special confining potential only recently by S.~Stringari
\cite{Str96} (see also the review articles
\cite{ParWal98,DalGioPitStr99,Griff99}), after the first experimental
successes of realizing Bose-Einstein condensation with trapped atomic vapors
\cite{Cor95,Ket95,Hul95}. This quantum hydrodynamic equation, shortly called
`Stringari equation', has a quite similar structure to the ordinary three
dimensional Schr\"odinger equation. The motive of the present investigation is
provided by the fact that similarly to the ordinary Schr\"odinger equation,
the solution of the Stringari equation is also only in very rare, special
cases available in analytic form
\cite{Str96,FliCsoSzeGra97,OhbSTWSh97,CsoRo98}. For this reason it is
important to see how the semiclassical methods developed for the ordinary
Schr\"odinger equation can be transported to the present problem, which has a
turning point structure of unusual type. In this work we demonstrate it in the
simplest, also analytically treatable case, where the confining potential is
spherically symmetric and harmonic. A comparison to the exact eigenvalues
shows that the accuracy of the semiclassical method is rather good, the
deviation is only a small shift in the energy square.

In the first section the Stringari equation is introduced, rewritten in
dimensionless form, separated in spherical polar coordinates, and the
analytic results are revisited.

In the second section we summarize briefly all the necessary information we 
need to know about the JWKB method for the present purposes, and also derive
how the form of a second order linear differential equation (containing no
term of first order derivative) changes under a general transformation of the
independent variable. This result will be used in the next section to
eliminate an undesirable term from the radial Stringari equation.

In the third section, after applying the above mentioned transformation the
semiclassical solution and the semiclassical quantization condition are
derived rigorously.

Finally the semiclassical results are compared to the exact analytic
ones, and some further possible applications of the JWKB method are
proposed.

\section{The Stringari equation --- analytic results}\label{sSe}
Let us consider a dilute Bose gas trapped in a confining potential
$V({\bf r})$. If the interparticle interaction is dominated by weak $s$-wave scattering, i.e., the (magnitude) of the $s$-wave scattering length $a$ is much smaller than the average distance between the particles,
and the temperature $T\ll T_{\rm c}$ is much less than the condensation temperature $T_{\rm c}$, i.e., the effect of the thermal cloud on the condensate is negligible then the `condensate wave
function' $\Psi({\bf r},t)=\langle\hat{\Psi}({\bf r},t)\rangle$, which
is the expectation value of the field operator $\hat{\Psi}({\bf
  r},t)$, satisfies a nonlinear partial differential equation, the so
called {\it Gross-Pitaevskii} equation \cite{Gro61,Gro63,Pit61}:
\begin{subequations}\label{eGP0}
\begin{equation}\label{eGP}
i\hbar\frac{\partial\Psi({\bf r},t)}{\partial t}=
\left(-\frac{\hbar^2}{2M}\Delta +V({\bf r})+g \abs{\Psi({\bf
r},t)}^2\right) \Psi({\bf r},t),
\end{equation}
where $M$ is the mass of the particles, and $g =\frac{4\pi \hbar^2 a}{M}$ is a
constant depending on the scattering length $a$. The validity conditions of the equation are 
\begin{equation}\label{evalGP}
LN^{-\frac{1}{3}} \gg \abs{a} \qquad \text{and} \qquad T \ll T_{\rm c},
\end{equation}
\end{subequations} 
where $L$ is the linear size of the condensate, and $N$ is the number of condensed atoms. In this paper we
restrict our attention to the $a>0$ (i.e., $g>0$) case, which means that the
interaction is repulsive.

In the hydrodynamic formalism two real quantities are introduced instead of the complex wave function of the condensate $\Psi({\bf r},t)= \sqrt{\varrho({\bf r},t)}e^{i\varphi({\bf r},t)}$, which are the condensate density $\varrho({\bf r},t) =\Psi \Psi^*$ describing the magnitude of the complex wave function and the velocity field ${\bf v}({\bf r},t) =\frac {i\hbar} {2M\varrho} (\Psi \nabla \Psi^* -\Psi^* \nabla \Psi ) =\frac {\hbar}{M} \nabla \varphi({\bf r},t)$ related to the phase of the condensate wave function. (For the sake of simplicity the space and time arguments $\bf r$, $t$ of the functions are omitted, if there is no danger of confusion.) Taking the time derivative of the density $\varrho$ and using the time dependent Gross-Pitaevskii equation (\ref{eGP}) for substituting the time derivatives of $\Psi$ and $\Psi^*$ the following continuity equation is obtained:
\begin{equation}\label{erhocont}
\frac{\partial \varrho}{\partial t} +\nabla(\varrho {\bf v})=0.
\end{equation}

Inserting the formula $\Psi =\sqrt{\varrho} e^{i\varphi}$ into the Gross-Pitaevskii equation (\ref{eGP}), and separating the real part of it, we get
\begin{equation}\label{eGPre}
\hbar\frac{\partial \varphi}{\partial t} =\frac {\hbar^2}{2M\sqrt{\varrho}}
\Delta\sqrt{\varrho} -\frac{\hbar^2}{2M}(\nabla \varphi)^2 -V -g\varrho.
\end{equation}
Taking the gradient of this equation, and using the expression ${\bf v}=\frac{\hbar}{M} \nabla \varphi$ for the velocity field, we obtain that the equation describing the time evolution of the velocity field ${\bf v}({\bf r},t)$ is 
\begin{equation}\label{etvvv}
M\frac{\partial{\bf v}}{\partial t} +\nabla \Bigl(
V+g\varrho+\frac{M{\bf v}^2}{2} -\frac{\hbar^2}{2M\sqrt{\varrho}}
\Delta\sqrt{\varrho} \Bigr)=0.
\end{equation}

It is worth stressing that the quantum hydrodynamic equations (\ref{erhocont}) and (\ref{etvvv}) are equivalent with the time dependent Gross-Pitaevskii equation (\ref{eGP}), they do not involve any further approximation, and they are valid in the linear as well as in the nonlinear regimes.

The ground state solution of the hydrodynamic equations is characterized by the overall zero value of the velocity field ${\bf v}({\bf r},t)$, in which case the equation (\ref{etvvv}) reduces to the time independent Gross-Pitaevskii equation
\begin{equation}\label{eGP2}
\mu \sqrt{\varrho({\bf r})} =\Bigl( -\frac{\hbar^2}{2M}\Delta +V({\bf r})
+g\varrho({\bf r}) \Bigr) \sqrt{\varrho({\bf r})},
\end{equation}
where $\mu$ is the chemical potential of the system, and $\Psi({\bf r},t) =\sqrt{\varrho({\bf r})} \exp \bigl(-\frac{i}{\hbar} \mu t \bigr)$.

If the number $N$ of the atoms in the condensate is sufficiently large, i.e., $Na \gg L$ (where $a>0$ is the scattering length of the repulsive interaction between the atoms, and $L$ is the linear size of the condensate) than the kinetic energy term $\propto \frac{\hbar^2} {\sqrt{\varrho}} \Delta \sqrt{\varrho}$ is almost everywhere negligible with respect to the other terms in equation (\ref{etvvv}). (This approximation fails only in a narrow region at the boundary of the condensate where $\varrho \to 0$.) The hydrodynamic equations have the following simpler form in this limit:
\begin{subequations}\label{eTF}
\begin{eqnarray}\label{eTFr}
\frac{\partial \varrho}{\partial t} +\nabla(\varrho {\bf v})&=&0,
\\ \label{eTFv}
M\frac{\partial{\bf v}}{\partial t} +\nabla \Bigl(
V+g\varrho+\frac{M{\bf v}^2}{2} \Bigr)&=&0.
\end{eqnarray}
The validity conditions of the above applied Thomas-Fermi approximation, together with the initially imposed conditions (\ref{evalGP}) are 
\begin{equation}\label{evalTF}
Na \gg L \gg N^{\frac{1}{3}} a>0, \qquad \text{and} \qquad T\ll T_{\rm c}.
\end{equation}
\end{subequations}

It is worth noting that this approximation has a great relevance, since the validity conditions (\ref{evalTF}) can be well satisfied in experiments \cite{DalGioPitStr99}.

Stringari \cite{Str96} investigated the time-periodic collective
excitational solutions of the Thomas-Fermi hydrodynamic equations (\ref{eTF}) in the linear (low energy) regime. The Thomas-Fermi ground state $\varrho_0({\bf r})$ corresponds to the zero velocity field ${\bf v}_0({\bf r}) =0$, in which case the solution 
\begin{equation}\label{erho0}
\varrho_0 ({\bf r}) =\left\{
\begin{array}{ll}
g^{-1}\bigl(\mu-V({\bf r}) \bigr),& \text{if } \mu-V({\bf r}) >0 \\
0 & \text{otherwise}
\end{array}
\right.
\end{equation}
is obtained from the equation (\ref{eTFv}). Linearizing the equations (\ref{eTF}) around the ground state $\varrho_0$ and ${\bf v}_0$, (substituting $\varrho({\bf r},t) =\varrho_0 ({\bf r}) + \delta \varrho ({\bf r},t)$ and ${\bf v}({\bf r},t) ={\bf v}_0 ({\bf r}) + \delta {\bf v}({\bf r},t)$ and keeping the first order terms) the following equations are obtained for the density and velocity fluctuations $\delta \varrho ({\bf r},t)$ and $\delta {\bf v}({\bf r},t)$
\begin{subequations}\label{elinTF}
\begin{eqnarray}\label{elinTFr}
\frac{\partial \delta \varrho}{\partial t} +\nabla (\varrho_0 \delta{\bf v})
&= &0,
\\ \label{elinTFv}
M\frac{\partial \delta {\bf v}}{\partial t} +
g\nabla\delta\varrho &=& 0.
\end{eqnarray}
\end{subequations}
Taking the time derivative of (\ref{elinTFr}) and using (\ref{elinTFv}) to eliminate (the time derivative of) the velocity fluctuation $\delta {\bf v}$ we obtain the time dependent Stringari equation
\begin{equation}\label{eStrt}
M\frac{\partial^2 \delta\varrho}{\partial t^2} =
\nabla\bigl( (\mu-V) \nabla \delta\varrho \bigr),
\end{equation}
and using the Ansatz $\delta \varrho ({\bf r},t) =e^{-i\omega t} \tilde{\xi}({\bf r})$ for the time dependence of the density fluctuation $\delta \varrho$ we arrive to the time independent Stringari equation
\begin{equation}\label{eStr}
-\nabla\Bigl( \bigl(\mu-V({\bf r})\bigr)
\nabla \tilde{\xi}({\bf r})\Bigr)=M\omega^2 \tilde{\xi}({\bf r}),
\end{equation}
where $\omega$ is the angular frequency of the excitation, and $\tilde{\xi}$ is defined only in the
region $\{{\bf r}|\mu-V({\bf r})>0\}$. (The `~$\tilde{}$~' (tilde) sign above
$\tilde{\xi}$ just designates that the variable is not dimensionless.) This equation is the starting point of our semiclassical investigations.

Let us suppose that the external trap potential function is isotropic.
In this case it is natural to rewrite equation (\ref{eStr}) in
spherical polar coordinates $\{r,\vartheta,\varphi\}$ and separate the
equation using the Ansatz $\tilde{\xi}
(r,\vartheta,\varphi)=\frac{\tilde{\chi}(r)}{r}
Y_{l,m}(\vartheta,\varphi)$. (As usual, $r=\abs{\bf r}$ and
$Y_{l,m}(\vartheta,\varphi)$ is the spherical harmonic function,
$l,m\in{\NN}$. Because of the denominator $r$ in the Ansatz, we
have $\int_{\RR^3} \abs{\tilde{\xi} ({\bf r})}^2 d^3r
=\int_{r=0}^\infty \abs{\tilde{\chi} (r)}^2 dr$.)  After separation,
one obtains the following second order ordinary differential equation
for the radial component $\tilde{\chi}(r)$:
\begin{multline}\label{er0}
U(r)\tilde{\chi}''(r)+U'(r)\tilde{\chi}'(r)+ \\
+\left( M\omega^2 -\frac
{U(r)}{r^2}l(l+1) -\frac{U'(r)}{r} \right) \tilde{\chi}(r)=0,
\end{multline}
where $U(r)$ stands for $\mu-V(r)$.

Let us rewrite the radial equation into dimensionless form! As length
unit it is reasonable to choose the Thomas-Fermi length $r_0$ which is
defined by the equation $V(r_0)=\mu$ i.e. $U(r_0)=0$, thus the new
independent radial variable is $\rho=\frac{r}{r_0} \in [0,1]$. For the
dimensionless energy (frequency) and potential we introduce the
variable $\varepsilon =\sqrt{ \frac{M}{2\mu}}r_0 \omega$ and the
function $u(\rho)=\frac{U(r)}{\mu}$, respectively. Using these new
notations, the equation (\ref{er0}) for the dimensionless radial
function $\chi(\rho)=\tilde{\chi}(r) =\tilde{\chi}(r_0 \rho)$ has the
form
\begin{multline}\label{er0.5}
u(\rho)\chi''(\rho)+u'(\rho)\chi'(\rho)+\\
+\left( 2\varepsilon^2-
\frac{u(\rho)}{\rho^2} l(l+1)-\frac{u'(\rho)}{\rho} \right) \chi(\rho) =0.
\end{multline}

Further we are interested in the case when the external trapping
potential is harmonic oscillator potential. This assumption makes the
problem not only analytically solvable
\cite{Str96,FliCsoSzeGra97,OhbSTWSh97,CsoRo98} but also has
experimental relevance \cite{Ketterle1999a,Cornell1999a}. For the sake
of simplicity, in this paper we restrict our attention to isotropic
harmonic oscillator potential $V({\bf r})=\frac{1}{2}M \omega_0^2
r^2$. It means that the Thomas-Fermi length is $r_0=\sqrt{\frac{2\mu}
  {M\omega_0^2}}$, for the dimensionless energy
$\varepsilon=\frac{\omega}{\omega_0}$ holds, and the dimensionless
potential function becomes $u(\rho)=1-\rho^2$. In this case the
equation~(\ref{er0.5}) takes the form
\begin{subequations}\label{er1}
\begin{multline}\label{er1a}
\chi''(\rho) -\frac{2\rho}{1-\rho^2} \chi'(\rho) +\\
+\left(
\frac{2(\varepsilon^2+1)}{1-\rho^2} -\frac{l(l+1)}{\rho^2} \right)
\chi(\rho) =0,
\qquad\rho \in [0,1]
\end{multline}
with the normalization and continuity condition
\begin{equation}\label{er1b}
\int_{\rho=0}^1 \abs{\chi(\rho)}^2 d\rho < \infty,\qquad
\begin{array}{ll}
{\displaystyle \lim_{\rho\searrow 0}} \left(\frac{\chi(\rho)}{\rho}\right)'
=0,& \text{if $l=0$,}\\
{\displaystyle \lim_{\rho\searrow 0}} \frac{\chi(\rho)}{\rho} =0,& 
\text{if $l\neq 0$.}
\end{array}
\end{equation}
\end{subequations}

This equation is an ordinary second order linear homogeneous differential
equation with three regular singular points at $0$, $1$ and $\infty$. It is
easy to see that in the Frobenius series expansion $\chi(\rho) =\rho^\alpha
\sum_{n\in\NN} \chi_n \rho^n$ of the solution $\chi(\rho)$ only the
coefficients $\chi_n$ of even indices $n$ are nonzero, thus the
equation~(\ref{er1}) can be further simplified by the substitution
$t=\rho^2$, $\kappa(t)=\chi(\rho)$ of the independent variable:
\begin{subequations}\label{erkap}
\begin{multline}\label{erkapa}
\kappa''(t)+\left( \frac{1}{2t} -\frac{1}{1-t} \right) \kappa'(t) +\\
+ \left(
\frac{\varepsilon^2+1}{2t(1-t)} -\frac{l(l+1)}{4t^2} \right) \kappa(t) =0,
\qquad t\in [0,1]
\end{multline}
\begin{equation} \label{erkapb}
\int_{t=0}^1 \frac{\abs{\kappa(t)}^2}{\sqrt{t}}  dt <\infty,
\qquad
\begin{array}{ll}
{\displaystyle \lim_{t \searrow 0}}
\frac{2t\kappa'(t)-\kappa(t)}{t} =0, & \text{if $l=0$,}\\
{\displaystyle \lim_{t \searrow 0}} \frac{\kappa(t)}{\sqrt{t}} =0,&
\text{if $l\neq 0$.}
\end{array}
\end{equation}
\end{subequations}

This equation has again three regular singularities at $t^{(0)} =0$,
$t^{(1)} =1$ and $t^{(\infty)} =\infty$ with indices $\alpha^{(0)}_1 =
\frac{l+1}{2}$, $\alpha^{(0)}_2 = -\frac{l}{2}$; $\alpha^{(1)}_1 =
\alpha^{(1)}_2 = 0$; and $\alpha^{(\infty)}_{1,2} = -\frac{1}{4}\pm
\frac{1}{2}\sqrt{2(\varepsilon^2 +1) +(l+ 1/2)^2} $, respectively,
what means that the basic solutions of the equation (\ref{erkapa})
have power-law behavior $\kappa(t) \propto t^{\alpha_{1,2}^{(0)}}$ at
the origin $t=0$, and one of them is locally linear (i.e.,
$\kappa(1)<\infty$, $\kappa'(1) < \infty$), and the other possesses
logarithmic singularity at $t= 1$. Because of the conditions
(\ref{erkapb}) the physically meaningful solutions must have the index
$\alpha_1^{(0)} =\frac{l+1}{2}$ at the origin and they should not have
logarithmic singularity at $t =1$. Inserting the Frobenius
series expansion
\begin{subequations}\label{ekap}
\begin{equation}\label{eFrs}
\kappa(t) = t^{\frac{l+1}{2}} \sum_{n=0}^\infty \kappa_n t^n,
\qquad \kappa_0 \ne 0
\end{equation}
into the differential equation (\ref{erkapa}), the recursion
\begin{equation}\label{ekrecn1}
(n+1)\big(n+l+\frac{3}{2}\big) \kappa_{n+1} =
\left( n^2+\frac{3}{2} n +nl +\frac{l}{2}-
\frac{\varepsilon^2}{2}\right) \kappa_n
\end{equation}
\end{subequations}
is obtained for the coefficients $\kappa_n$, with the initial condition
$\kappa_0 \ne 0$. The regularity condition of $\kappa(t)$ at $t =1$ can be
satisfied by requiring the expansion (\ref{eFrs}) to be finite, i.e., that
$\kappa_k =0$ for all $k\ge n_0$, what results in the discrete energy spectrum
\cite{Str96} 
\begin{equation}\label{espec}
\varepsilon(n,l)=\sqrt{2n^2 +2nl +3n +l},\qquad n,l \in \NN.
\end{equation}

\begin{figure*}
\centerline{\scalebox{0.8}{\input{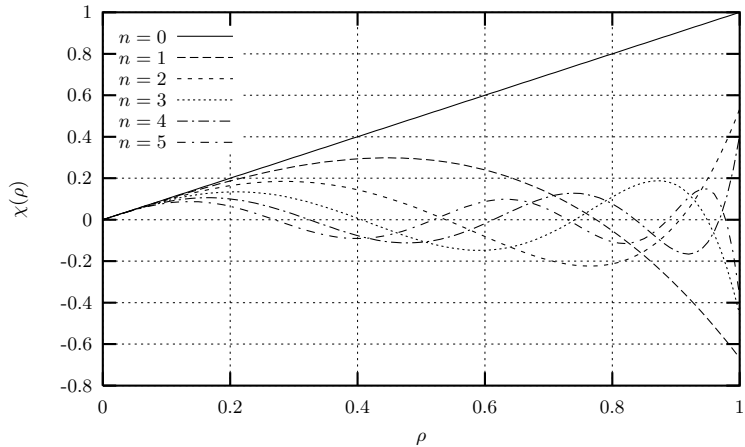}}}
\caption{\label{f:n}The exact (dimensionless) radial wave function $\chi(\rho)$ for quantum numbers $n=0,1,2,3,4,5$ and $l=0$.}
\end{figure*}

The figures \ref{f:n} and \ref{f:l} show the exact radial wave
function $\chi(\rho)$ for a few quantum numbers.

\begin{figure*}
\centerline{\scalebox{0.8}{\input{cr3l}}}
\caption{\label{f:l}The exact (dimensionless) radial wave function $\chi(\rho)$ for quantum numbers $n=3$ and $l=0,1,2,3$.}
\end{figure*}

It is worth noting that by changing the dependent variable according
to the formula $\kappa(t) =t^{\alpha^{(0)}_1} w(t)$ the equation can
be transformed to the hypergeometric form \cite{AbrSte}
\begin{subequations}
\begin{equation}\label{ehypg}
t(1-t)\frac{d^2 w}{dt^2} +\big( c-(a+b+1)t\big) \frac{dw}{dt} -abw =0
\end{equation}
with parameter values 
\begin{eqnarray}\nonumber
ab&=& -\frac{\varepsilon^2 -l}{2},\qquad a+b=c= l+\frac{3}{2},\\ 
\label{eabc}
a,b&=& \frac{l+\frac{3}{2} \pm \sqrt{l^2 +l+\frac{9}{4} +2\varepsilon^2}}{2},
\end{eqnarray}
and additional physical constrains
\begin{equation}\label{ewcon}
\begin{array}{c}
{\displaystyle \int_{t=0}^1 t^{l+1/2} \abs{w(t)}^2 dt <\infty,} \\
\begin{array}{ll}
{\displaystyle \lim_{t \searrow 0}} \big(w(t)+2t w'(t)\big) t^{\frac{l-1}{2}}
=0, & \text{if $l=0$,}\\
{\displaystyle \lim_{t \searrow 0}} w(t)t^{\frac{l-1}{2}} =0,&
\text{if $l\neq 0$.}
\end{array}
\end{array}
\end{equation}
\end{subequations}
The physically meaningful solutions [satisfying (\ref{ewcon})] fall
into the so called {\it degenerate case} of the hypergeometric
equation, in which case the solution has the form $w(t)=t^\alpha
(1-t)^\beta p_n (t)$ where $p_n (t)$ is a finite polynomial of degree
$n$. The condition for this case is that either $a=c-b$ or $b=c-a$ is
an integer, and it reproduces the exact quantization condition
(\ref{espec}).

\section{Preliminary steps}
In this section first the necessary formulae of the JWKB method are
summarized, and then, for later use, we briefly derive how the
equation subjected to JWKB approximation transforms under a general
change of the independent variable.

\subsection{The JWKB semiclassical method}\label{ssJWKB}
The JWKB method, originally developed by H.\ Jeffreys \cite{Jef25}, G.\
Wentzel \cite{Wen26}, H.\ A.\ Kramers \cite{Kra26} and L.\ Brillouin
\cite{Bri26a,Bri26b} has still remained one of the most powerful tools
for the approximative solution of wave equations since its birth in
the very early days of quantum mechanics
\cite{FroFro65,BerMou72,OrsBen78,Chi91}. Basically it can be used for
obtaining a global approximation to the solution of a linear
differential equation whose highest derivative is multiplied by a
small parameter, and the method gives an asymptotic approximation of
the real solution on a certain interval in terms of increasing powers
of this parameter.

In this paper we consider only second order homogeneous linear
differential equations of the following type:
\begin{equation}\label{esc0}
  \delta^2 y''(x)+ Q(x) y(x) =0,
\end{equation}
where $\delta^2$ is the small real parameter. In this case the JWKB
Ansatz has the form
\begin{equation}\label{eAns}
y(x)=\exp \left(\frac{1}{\delta} S_\delta (x)\right),\qquad
S_\delta (x) = \sum_{n=0}^\infty \delta^n S_n (x).
\end{equation}

Substituting this Ansatz into the differential equation~(\ref{esc0})
and requiring the equivalence in any orders of $\delta$ the following
equations are obtained for the multiplier functions $S_n$ in the
exponent \cite{OrsBen78}:
\begin{subequations}\label{escS}
\begin{align}\label{escS0}
S_0 (x)&=\pm i \int_{t=t_0}^x \sqrt{Q(t)} dt,
\\ \label{escS1}
S_1 (x)&=-\frac{1}{4} \ln Q(x),
\\ \label{escS2}
S_2 (x)&= \pm i \int_{t=t_0}^x \left( \frac{5{Q'}^2 (t)}{32Q^{5/2} (t)} -
\frac{Q''(t)}{8Q^{3/2}(t)}  \right) dt,
\quad \text{etc}
\end{align}
\end{subequations}
thus in the first order the semiclassical solution has the general
form
\begin{widetext}
\begin{eqnarray}\nonumber
y_1 (x) &=& C_1\exp\Big(\frac{S_0(x)}{\delta}+S_1 (x)\Big) +
 C_2\exp\Big(-\frac{S_0(x)}{\delta}+S_1 (x)\Big) =
\\ \label{escsol}
&=&C_1 Q^{-1/4} 
\exp\left( \frac{i}{\delta} \int_{t=t_0}^x \sqrt{Q(t)} dt \right)+
C_2 Q^{-1/4} 
\exp\left( -\frac{i}{\delta} \int_{t=t_0}^x \sqrt{Q(t)} dt \right),
\qquad C_1, C_2 \in \RR.
\end{eqnarray}
\end{widetext}

The JWKB solution, truncated at the $N$-th order $y_N (x)=\exp\big(
\frac{1}{\delta} \sum_{n=0}^N \delta^n S_n (x) \big)$ is a uniformly valid
approximation on the interval $I\subset \RR$ in the $\delta \to 0$
limit, if the following relations \cite{OrsBen78} hold for the succeeding
terms in the exponent:
\begin{subequations}\label{esccond}
\begin{align}\label{escconda}
\frac{1}{\delta}S_0(x)&\gg S_1(x) \gg \delta S_2(x) \gg \dots 
\gg \delta^{N-1} S_N(x),
\\ \label{esccondb}
1&\gg \delta^N S_{N+1} (x), 
\qquad \text{for}\quad \delta \to 0, \quad x\in I.
\end{align}
\end{subequations}

In concrete cases these asymptotic inequalities define the interval
$I$ on which the JWKB approximation can be applied. On the regions
where the approximation fails (in most cases at the classical
turning points, where $Q(x)=0$), usually the multiplier function
$Q(x)$ itself should be approximated by a simpler function, for which
the equation~(\ref{esc0}) is exactly solvable, and then the
semiclassical solution (of the exact equation) and the exact solution
(of the approximate equation) should be patched together. This
procedure will be carried out in detail for the radial Stringari
equation, (which is an equation with unusual types of turning points)
in the \ref{sssrS}$^{\text{th}}$ Section.

\subsection{The effect of the change of the independent variable}\label{ssec}
It is well known that the applicability of the JWKB method highly
depends on the proper choice (transformation) of the independent
variable of the differential equation investigated. For this reason
the formulae describing the transformation of the differential
equation induced by a general change of the independent variable are
stated here.

The starting point is the equation~(\ref{esc0}). Let us transform the
independent variable $x$ in the equation to the new variable $t$ by
the invertible transformation
\begin{subequations}\label{e:tr}
\begin{eqnarray}
\X &:&\RR\rightarrowtail\RR,\qquad\quad
\makebox[1em][r]{$t$} \mapsto \X(t),
\\
\T=\X^{-1} &:&\RR\rightarrowtail\RR,\qquad\quad
\makebox[1em][r]{$x$} \mapsto \T(x),
\end{eqnarray}
\end{subequations}
which is defined on a suitable interval of the real line. Denoting the
new dependent variable by $\bar{y}(t)=y(\X(t))$, the original
equation~(\ref{esc0}) has the transformed form
\begin{subequations}\label{e:1}
\begin{equation}\label{e:1_a}
\delta^2\bar{y}''(t) +\delta^2\bar{P}(t) \bar{y}' (t) 
+\bar{Q} (t)\bar{y}(t) =0,
\end{equation}
where the multiplier functions are
\begin{eqnarray}\nonumber
\bar{P}&=&-\frac{\X''}{\X'} =-\big(\ln (\X')\big)' 
=\frac{\T''}{{\T'}^2}\circ\X= \big(\ln(\T')\circ \X\big)',
\\ \label{e:PQ1}
\bar{Q} &=&(Q \circ \X){\X'}^2= \left( \frac{Q}{{\T'}^2} \right) \circ \X.
\end{eqnarray}
\end{subequations}

The term containing the first order derivative $\bar{y}'(t)$ can be
eliminated by applying the
\begin{equation}\label{etr1d}
\tilde{y}(t)=\exp \left( \frac{1}{2} \int_{z=z_0}^t
\bar{P}(z)dz\right) \bar{y}(t) =\frac{\bar{y}(t)} {\sqrt{\X'(t)}}
\end{equation}
transformation of the dependent variable. After the transformation the
equation
\begin{subequations}\label{e:2}
\begin{equation}\label{e:2e}
\delta^2\tilde{y}''(t)+\tilde{Q}(t)\tilde{y}(t)=0
\end{equation}
is obtained, where the new coefficient function is 
\begin{eqnarray}\nonumber
\tilde{Q}&=&(Q \circ \X){\X'}^2 +
\delta^2\frac{2\X'''\X'- 3{\X''}^2}{4{\X'}^2}=
\\ \label{e:Q2}
&=&\bigg(\frac{1}{{\T'}^2} \Big( Q +\delta^2
\big(-\frac{1}{2} \Pi' +\frac{1}{4}\Pi^2 \big)\Big) \bigg) \circ \X
\end{eqnarray}
with
\begin{equation}
\Pi = \frac{\T''}{\T'} =(\ln \T')', 
\end{equation}
\end{subequations}
and the new unknown function $\tilde{y}(t)$ is related to the original
unknown function $y(x)$ via the formula
\begin{equation}\label{e:yty}
\tilde{y}=\frac{1}{\sqrt{\X'}}(y\circ\X) =\left( \sqrt{\T'} y\right) \circ \X.
\end{equation}

Thus applying two consecutive transformations to the differential
equation (\ref{esc0}), first an arbitrary invertible change of the
independent variable (\ref{e:tr}) and then an appropriate
transformation (\ref{etr1d}) of the dependent variable, one arrives to
the equation~(\ref{e:2e}) of the same form, but with different
coefficient function~(\ref{e:Q2}). This freedom of the choice of the
independent variable enables us to transform the investigated equation
to a more appropriate form before applying the JWKB method.

\section{The semiclassical solution of the radial Stringari
equation (In case of $\lowercase{l}=0$)}\label{sssrS}
In this section the JWKB method, summarized in the previous section is
applied to the (dimensionless radial) Stringari equation (\ref{er1})
derived in Section \ref{sSe}.

As we did in the previous subsection [formula (\ref{etr1d})], we can
get rid of the term containing the first order derivative
$\chi'(\rho)$ in equation (\ref{er1a}) by applying the transformation
$y(\rho):= e^{-\int_{z=0}^\rho \frac{z}{1-z^2} dz} \chi(\rho)
=\sqrt{1-\rho^2}\chi(\rho)$. As result, the equation
\begin{subequations}\label{e:St1+}
\begin{equation}\label{e:St1}
\delta^2 y''(x)+\big( Q (x) +\delta^2 W_l (x)\big) y(x)=0,
\qquad x\in [0,1]
\end{equation}
\begin{equation}\label{e:St1Q} 
Q (x) =\frac{1}{1-x^2},\qquad
W_l (x) = -\frac{L}{x^2}+ \frac{1}{(1-x^2)^2} 
\end{equation}
is obtained with the normalization and continuity conditions
\begin{equation}\label{e:St1_}
\int_0^1 \frac{y^2(x)}{1-x^2}dx <\infty,\qquad
\begin{array}{ll}
{\displaystyle \lim_{x\searrow 0}} \left(\frac{y(x)}{x}\right)' =0,&
\text{if $l=0$,}\\
{\displaystyle \lim_{x\searrow 0}} \frac{y(x)}{x} =0,&
\text{if $l\neq 0$,}
\end{array}
\end{equation}
where the abbreviations 
\begin{equation}\label{e:EL}
\delta^2 =\frac{1}{2\varepsilon^2+2},\qquad\quad L=l(l+1)
\end{equation}
\end{subequations}
were used. (For convenience we denoted here the independent variable
with $x$ instead of $\rho$.) This equation is the starting point of
the semiclassical quantization, and $\delta^2$ is used as small
parameter.

A comparison of the equation (\ref{e:St1}) to the general form
(\ref{esc0}) shows that (\ref{e:St1}) does not have the desired form
since it contains a small term $\delta^2 W_l (x)y(x)$.

First, to eliminate the $W_0 (x)= \frac{1}{(1-x^2)^2}$ part of this
term, (which does not depend on the angular momentum quantum number
$l$), we apply a transformation (\ref{e:tr}--\ref{etr1d}), discussed
in Subsection \ref{ssec}, to the equation (\ref{e:St1+}), and
choose the transforming function $\T(x)$ to be a solution of the
Ricatti-equation
\begin{equation}\label{eRic}
\frac{1}{2}\Pi'(x) -\frac{1}{4}\Pi^2 (x)=W_0(x)=\frac{1}{(1-x^2)^2},
\end{equation}
where $\Pi=\frac{\T''}{\T'} =(\ln \T')'$. It is easy to see that 
\begin{equation}\label{ePTX}
\Pi(x)= \frac{2x}{1-x^2},\quad \T(x)= \arth(x),\quad \X(t)= \tanh(t)
\end{equation}
is a simple solution of the nonlinear differential equation
(\ref{eRic}), and for this choice the transformed equation (\ref{e:2})
has the form
\begin{subequations}\label{e:yt}
\begin{equation}\label{e:yta}
\delta^2 \tilde{y}''(t)+ 
\big(\tilde{Q}(t)+\delta^2\tilde{W}_l (t)\big) \tilde{y}(t)=0,\quad
t\in [0,\infty)
\end{equation}
with
\begin{equation} \label{e:ytb}
\tilde{Q} (t) =\frac{1}{\cosh^2 (t)},\qquad
\tilde{W}_l (t) =\frac{-L}{\sinh^2 (t)}
\end{equation}
for the transformed variable $\tilde{y} (t)=\cosh(t)
y\big(\tanh(t)\big)$. Furthermore, the physically relevant solutions
must obey the supplementary conditions
\begin{equation}\label{e:ytcon}
\int_{t=0}^\infty \frac{\tilde{y}^2(t)}{\cosh^2 (t)} dt <\infty, \qquad
\begin{array}{ll}
{\displaystyle \lim_{t\searrow 0}} \left(\frac{\tilde{y} (t)}{t}\right)' =0,&
\text{if $l=0$,}\\
{\displaystyle \lim_{t\searrow 0}} \frac{\tilde{y} (t)}{t} =0,&
\text{if $l\neq 0$.}
\end{array}
\end{equation}
\end{subequations}

We note that the correspondence
\begin{equation}\label{e:chity}
\chi\bigl(\tanh(t)\bigr) =\tilde{y}(t)
\end{equation}
holds between the original variable $\chi(\rho)$ of the equation
(\ref{er1a}) and the transformed variable $\tilde{y}(t)$ of
(\ref{e:yta}).

To get rid of the undesirable term $\delta^2 \tilde{W}_l$ in the
followings we restrict our attention to the $l=L=0$ case, where this
problem does not occur, since $\tilde{W}_0(t) =0$.

\begin{figure*}
\center{\scalebox{0.8}{\input{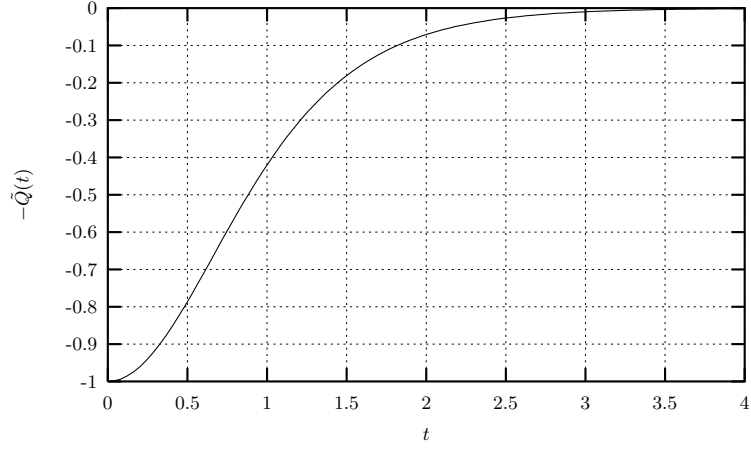}}}
\caption{\label{f:Q}The dimensionless potential function $-\tilde{Q}(t) 
=\frac{-1}{\cosh^2(t)}$.}
\end{figure*}

In this case the differential equation~(\ref{e:yta}) has the simpler form
\begin{subequations}\label{e:yt0}
\begin{equation}\label{e:yt0a}
\delta^2 \tilde{y}''(t)+\tilde{Q} (t) \tilde{y}(t) = 0 \qquad\quad
\text{with}
\end{equation}
\begin{equation} \label{e:yt0b}
\tilde{Q} (t)= \frac{1}{\cosh^2 (t)},\qquad\quad
\delta=\frac{1}{\sqrt{2\varepsilon^2 +2}},
\end{equation}
\end{subequations}
where the dimensionless potential function $-\tilde{Q}(t)$, depicted
in figure~\ref{f:Q} has a quite unusual form. There are no real
turning points, the function $-\tilde{Q}(t)$ builds a parabolic
potential well around the origin (for $0\le t\ll1$), and it approaches
exponentially to zero from below as $t\to\infty$.

According to the formulae~(\ref{escS}) in Subsection \ref{ssJWKB},
the action variables $S_n (t)$ (and their asymptotic behavior) in the
semiclassical solution of equation~(\ref{e:yt0}) are
\begin{subequations}\label{e:S}
\begin{align}\nonumber
S_0(t)&= \pm 2i \left( \arctan(e^t) -\frac{\pi}{2} \right)
\\ \label{e:S_a}
&\sim \makebox[4.5cm][l]{${\displaystyle \mp 2i e^{-t}}$} (t\gg 1),
\\ \nonumber
S_1(t)&= \frac{1}{2}\ln \bigl( \cosh(t)\bigr) -\frac{i\pi}{4}
\\ \label{e:S_b}
&\sim \makebox[4.5cm][l]{${\displaystyle \frac{t}{2}}$} (t\gg 1),
\\ \label{e:S_c}
S_2(t)&\sim \makebox[4.5cm][l]{${\displaystyle \pm \frac{i}{16} e^t}$} (t\gg
1), 
\end{align}
\end{subequations}
Applying the formula (\ref{escsol}) and the validity criterion
(\ref{esccond}), (which has the form $\delta e^t \ll 1$ in the
present special case), the following first order semiclassical
solution is obtained for the equation (\ref{e:yt0}):
\begin{subequations}\label{e:ysc}
\begin{align}\nonumber
\tilde{y}_{\text I} (t)&=
A_{\text I}\sqrt{\cosh(t)}\sin\left(\frac{2}{\delta}\left(
\arctan (e^t) -\frac{\pi}{4}\right) \right) 
\\ \label{e:ysc_a}
&\hspace{3cm} \text{if } 0\le t\ll -\ln(\delta),
\\ \nonumber
&\sim 
\frac{A_{\text I}}{\sqrt{2}}
e^{t/2} \sin\left( \frac{2}{\delta}\left( \frac{\pi}{4}-e^{-t}
\right) \right) 
\\ \label{e:ysc_b}
&\hspace{3cm} \text{if } 1\ll t \ll -\ln(\delta),
\end{align}
\end{subequations}
\noindent
where $A_{\text I}\in \RR$. (In the $t\gg 1$ approximation the
estimation $\frac{\pi} {2} -\arctan x =\int_{u=x}^\infty
\frac{du}{1+u^2} \approx \int_{u=x}^\infty \frac{du}{u^2}
=\frac{1}{x}$ was used inside the argument of the sine function, which is
valid in the $x=e^t \gg1$ limit.)

To get a good approximative solution of the equation~(\ref{e:yt0}) for
large $t\gg 1$ values it is straightforward to approximate the
coefficient function by $\tilde{Q}_0 (t) =\frac{1}{\cosh^2 (t)} \sim 4
e^{-2t}$ if $1\ll t$. With this approximation, however, the obtained
differential equation $\delta^2 \tilde{y}_{\text{II}}''(t)+4 e^{-2t}
\tilde{y}_{\text{II}} (t) =0$ is exactly solvable; the two basic solutions
$\tilde{y}_{\text{II}}$, $\tilde{y}_{\text{II}}^\dag$ are
\begin{equation}\label{e:yII}
\begin{array}{ll}
{\displaystyle 
\tilde{y}_{\text{II}}(t) = A_{\text{II}} 
J_0 \left(\frac{2}{\delta} e^{-t}\right) }
& \qquad\text{and} \\
{\displaystyle 
\tilde{y}_{\text{II}}^\dag (t) =
A_{\text{II}}^\dag Y_0 \left( \frac{2}{\delta} e^{-t}\right), }
& \qquad\text{if } 1\ll t,
\end{array}
\end{equation}
where $J_0$ and $Y_0$ are the zeroth order Bessel functions of the
first and second kind, respectively \cite{AbrSte}, and $A_{\text{II}},
A_{\text{II}}^\dag \in \RR$. The function $\tilde{y}_{\text{II}}^\dag (t)$ is
linearly diverging for $t \to \infty$ values what, according to the
correspondence~(\ref{e:chity}), causes the divergence of $\chi(\rho)$ at
$\rho\to 1$, so the physically meaningful solution of our problem is only
$\tilde{y}_{\text{II}} (t)$.

\begin{figure*}
\center{\includegraphics[scale=0.35]{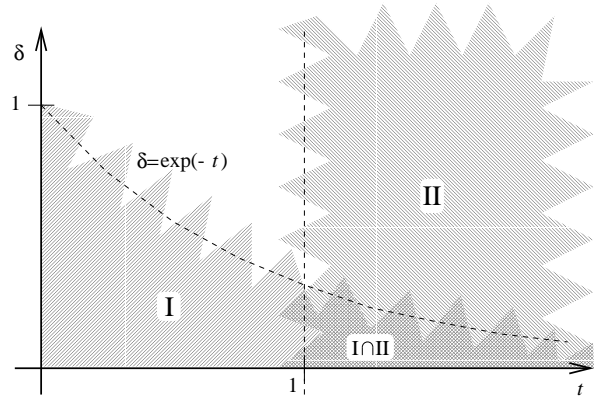}}
\caption{\label{f:int} The validity regions of the two approximative solutions
  $\tilde{y}_{\text{I}} (t)$ [equation~(\ref{e:ysc})] and
  $\tilde{y}_{\text{II}} (t)$ [equation~(\ref{e:yII})] on the
  $t-\delta$ plane.}
\end{figure*}

Thus we have obtained two approximative solutions for the
equation~(\ref{e:yt0}), which are valid in two different, but
especially for small $\delta \to 0$ values exceedingly overlapping
intervals, as it is shown in figure~\ref{f:int}. To get a
semiclassical quantization condition the functions
$\tilde{y}_{\text{I}} (t)$ [equation~(\ref{e:ysc})] and
$\tilde{y}_{\text{II}} (t)$ [equation~(\ref{e:yII})] should be matched
in the overlapping region $1\ll t \ll -\ln(\delta)$. At first sight
the two functions bear no readily visible resemblance of form,
however, using the asymptotic expansion $J_0 (z) \sim
\sqrt{\frac{2}{\pi z}}\cos \left(z-\frac{\pi}{4}\right)$ for the
Bessel function, which is valid for $|z|\gg 1$ values \cite{AbrSte},
the approximation
\begin{multline}\label{e:yIIas}
\tilde{y}_{\text{II}} (t) \sim \sqrt{\frac{\delta}{\pi}} 
A_{\text{II}} e^{t/2}
\cos\left(\frac{2}{\delta} e^{-t} -\frac{\pi}{4} \right),
\\ 
\text{if } 1 \ll t \ll -\ln(\delta)
\end{multline}
is obtained, which can well be compared to the
formula~(\ref{e:ysc_b}). The matching condition between the two
asymptotic formulae (\ref{e:ysc_b}) and (\ref{e:yIIas}) is clearly the
requirements $A_{\text{II}} = \pm \sqrt{ \frac{\pi}{2\delta}} A_{\text
  I}$ and $\sin\left( \frac{2}{\delta} \left( \frac{\pi}{4}-e^{-t}
  \right) \right) =\pm\cos \left( \frac{2}{\delta} e^{-t}
  -\frac{\pi}{4} \right)$ for every $t$ value in the interval $1\ll
t\ll -\ln(\delta)$. This means that the sum of the arguments of the
trigonometric functions must be $n \pi +\pi /2$, where $n \in \NN$ is
a nonnegative integer.  Using the definition~(\ref{e:yt0b}) of
$\delta$ the following semiclassical energy spectrum is obtained
\begin{equation}\label{e:spsem}
\varepsilon_{\text{sc}} (n) =\sqrt{2n^2 +3n+\frac{1}{8}},
\qquad l=0,\enskip n \in \NN,
\end{equation}
which differs from the exact spectrum (\ref{espec}) only by a
constant shift of $\frac{1}{8}$ under the square root sign.

With the help of the formula (\ref{e:chity}) the semiclassical radial
wave functions $\tilde{y}_{\text{I}} (t)$ [equation (\ref{e:ysc_a})] and
$\tilde{y}_{\text{II}} (t)$ [equation (\ref{e:yII})] can be transformed
back to approximate the original radial function $\chi (\rho)$
[in equation (\ref{er1})]:
\begin{subequations}\label{e:chisc}
\begin{eqnarray}\nonumber
\chi_{\text{I}} (\rho) &=& \tilde{y}_{\text{I}} \big(\arth(\rho)\big) =
\delta (1-\rho^2)^{-\frac{1}{4}} \times
\\ \label{e:chisc_a}
&\times&  
\sin \bigg( \frac{2}{\delta} \Big(\arctan \sqrt{\frac{1+\rho}{1-\rho}} 
-\frac{\pi}{4}\Big) \bigg),
\\ \nonumber
\chi_{\text{II}} (\rho) &=& \tilde{y}_{\text{II}} \big(\arth(\rho)\big) =
\\ \label{e:chisc_b}
&=& (-1)^n \sqrt{ \frac{ \pi \delta} {2}} 
J_0 \bigg(\frac{2}{\delta} \sqrt{ \frac{1-\rho}{1+\rho}} \bigg).
\end{eqnarray}
\end{subequations}
(Here the `normalization' condition $\kappa_0 =\tilde{y}'(0)=
\chi'(0)= 1$ was used, which implies that the multiplier constants in
the semiclassical formulae (\ref{e:ysc}) and (\ref{e:yII}) are
$A_{\text I} =\delta$ and $A_{\text{II}} =(-1)^n \sqrt{\frac{\pi
    \delta}{2}}$.)

\begin{figure*}
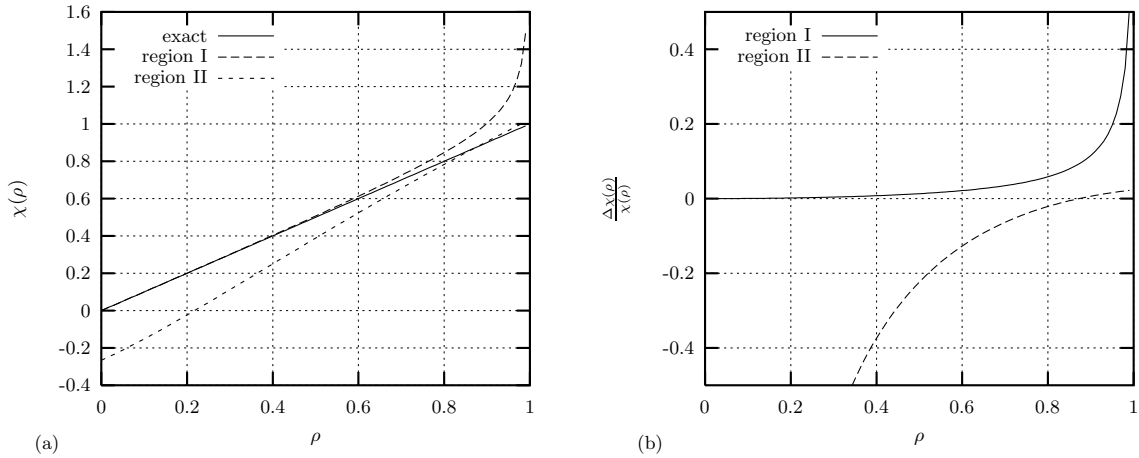

\center{
\begin{tabular}{ccc}
\scalebox{0.8}{\input{acc_00a}} &\hspace*{1cm}& \scalebox{0.8}{\input{acc_00r}}
\end{tabular}
}
\caption{\label{f:acc0}{\it a)} The exact and the semiclassical (dimensionless) radial wave
  functions [$\chi(\rho)$, $\chi_{\text{I}}(\rho)$ and
  $\chi_{\text{II}}(\rho)$]; {\it b)} The relative error
  $\frac{\Delta\chi(\rho)}{\chi(\rho)}$ of the semiclassical wave
  functions (where $\Delta\chi$ is $\chi_{\text{I}}-\chi$ or
  $\chi_{\text{II}} -\chi$) for radial quantum number $n=0$.}
\end{figure*}

\begin{figure*}
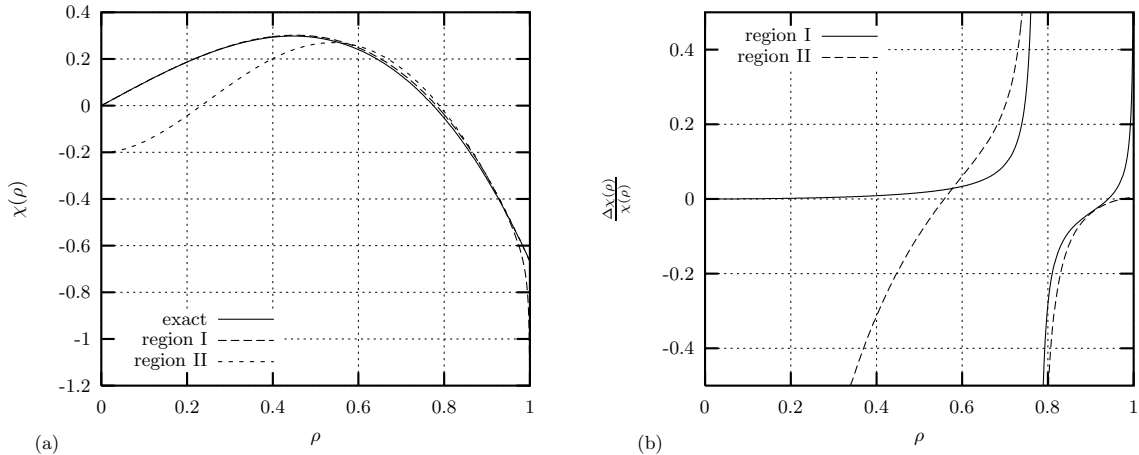

\center{
\begin{tabular}{ccc}
\scalebox{0.8}{\input{acc_01a}} &\hspace*{1cm}& \scalebox{0.8}{\input{acc_01r}}
\end{tabular}
}
\caption{\label{f:acc1}{\it a)} The exact and the semiclassical (dimensionless) radial wave
  functions [$\chi(\rho)$, $\chi_{\text{I}}(\rho)$ and
  $\chi_{\text{II}}(\rho)$]; {\it b)} The relative error
  $\frac{\Delta\chi(\rho)}{\chi(\rho)}$ of the semiclassical wave
  functions (where $\Delta\chi$ is $\chi_{\text{I}}-\chi$ or
  $\chi_{\text{II}} -\chi$) for radial quantum number $n=1$.}
\end{figure*}

To get an insight into the accuracy of the semiclassical
approximation, in figures \ref{f:acc0} and \ref{f:acc1} we plotted the
exact [$\chi( \rho) =\kappa( \rho^2)$, equation (\ref{ekap})] and the
semiclassical wave functions [$\chi_{\text I} (\rho)$,
$\chi_{\text{II}} (\rho)$ equations (\ref{e:chisc})] as well as the
relative errors $\frac {\chi_{\text{I}} (\rho) -\chi(\rho)}
{\chi(\rho)}$ and $\frac {\chi_{\text{II}} (\rho) -\chi(\rho)}
{\chi(\rho)}$ for the lowest two radial quantum numbers $n=0,1$. The
exact radial function was calculated with the exact energy eigenvalues
(\ref{espec}) [$l=0$] and with the `normalization' condition $\kappa_0
=\tilde{y}'(0)=1$, while the semiclassical functions $\chi_{\text I}
(\rho)$ and $\chi_{\text{II}} (\rho)$ were calculated with the
semiclassical energy spectrum (\ref{e:spsem}) (and with the same
`normalization' condition).

The figures \ref{f:acc0}, \ref{f:acc1} demonstrate that even for the
lowest energy levels the semiclassical wave function $\chi_{\text{I}}
(\rho)$ already approximates the exact function $\chi(\rho)$ fairly
well on almost the whole interval $[0,1]$. The only failure of
$\chi_{\text{I}} (\rho)$ is that it diverges at $\rho =1$, while the
exact function $\chi(\rho)$ has finite value with finite slope at this
point. This `illness' is cured by the second approximative solution
$\chi_{\text{II}} (\rho)$ which is quite different from the exact
function on almost the whole interval $[0,1]$, but in the immediate
vicinity of $\rho =1$, where $\chi_{\text{I}} (\rho)$ deviates from
$\chi(\rho)$, gives a good approximation.

We remark that the divergence of the relative error at the roots
$\rho_*$ of the exact function $\chi(\rho_*)=0$ are due to an
unavoidable, small shift between the roots of the exact and
semiclassical functions.

\begin{figure*}
\center{\scalebox{0.8}{\input{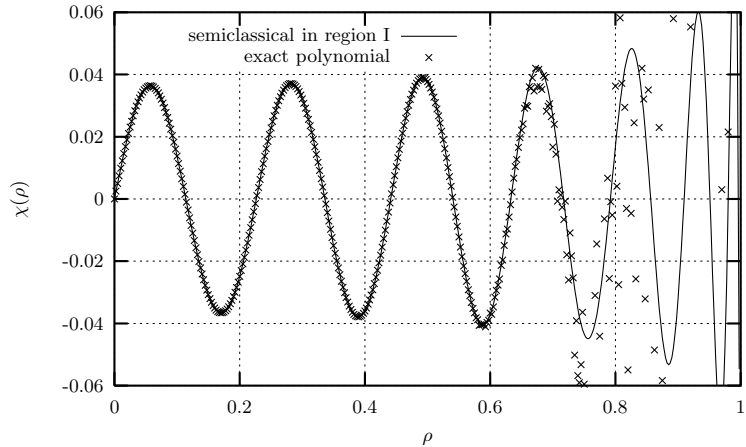}}}
\caption{\label{f:numstab} The semiclassical 
$\chi_{\text{I}} (\rho)$ and the exact $\chi(\rho)$ dimensionless wave functions 
calculated by standard $C$ routines for quantum numbers $l=0$ and $n=13$.}
\end{figure*}

In the end we find it worth making a final remark upon the numerical
stability of the calculation of the exact and semiclassical wave
function. Figure \ref{f:numstab} shows the semiclassical and
the exact wave function, $\chi_{\text{I}} (\rho)$ and $\chi(\rho)$ for
a somewhat larger but still relatively small radial quantum number
$n=13$, calculated with the help of standard $C$ routines with the
usual floating point accuracy. It is clearly visible that above a
certain point $\rho_0 \approx 0.7$ the exact polynomial totally looses
its numerical stability, while the semiclassical function can still be
easily calculated and serves as a very good approximation almost on
the whole interval (apart from the II$^{\text{nd}}$ region
$(0.997,1]$, where $\chi_{\text{II}}(\rho)$ becomes the more
appropriate approximation).  The table~\ref{t:numstab}
shows the (exact values of the) coefficients $\kappa_i$ of the
polynomial (\ref{eFrs}) obtained from the recursion (\ref{ekrecn1}) in
the $l=0$, $n=13$ case. The unexpectedly large values of these
coefficients cause the numerical instability in the calculation of the
exact function $\chi(\rho)$.  Therefore, in certain cases it may well
be reasonable to use semiclassical wave functions instead of the exact
ones in numerical calculations.

\begin{table}
\caption{\label{t:numstab} The coefficients $\kappa_i$ [equation 
(\ref{ekap})] for $l=0$ and $n=13$}
\begin{tabular}{||r|d||}
\hline\hline
$i$ & \kappa_i \\ \hline
$0$ & 1.\\
$1$ & -125.6667\\
$2$ & 4674.7998\\
$3$ & -80807.25\\
$4$ & 785626.0625\\
$5$ & -4756608.5\\
$6$ & 19026434.\\
$7$ & -52005588.\\
$8$ & 98657656.\\
$9$ & -129812704.\\
$10$ & 116213280.\\
$11$ & -67523128.\\
$12$ & 22957864.\\
$13$ & -3466572.\\
\hline\hline
\end{tabular}
\end{table}

\section{Discussion}
In the previous sections the JWKB semiclassical approximation has been
rigorously carried out for a very special case of the Stringari
equation (\ref{eStr}) (with spherically symmetric harmonic external
potential $V({\bf r})$ and $l=0$ zero angular momentum quantum
number). Because of the unusual type of the turning point, the
vicinity of this point needed special treatment and careful analysis.
A comparison of the semiclassical results to the known exact solutions
and eigenvalues shows that the accuracy of the semiclassical
approximation is quite satisfying; there is only a small
($\frac{1}{8}$ in dimensionless units) shift in the energy square, and
the relative error of the semiclassical radial wave function is
already for the lowest energy levels under a few percent. For larger
quantum numbers the semiclassical wave function is numerically much
more stable than the exact polynomial, what may make the use of the
semiclassical functions preferable than the exact ones in certain
numerical calculations.

We remark that a simpler, more practical way of achieving the same
first order semiclassical solution~(\ref{e:ysc_a}) of the radial
Stringari equation~(\ref{e:yt0}) in the $l=0$ case would have been
just to insert the Ansatz~(\ref{eAns}) right into the
equation~(\ref{e:St1}), without bothering us about the $\delta^2 W_0
(x)$ term.  This term is so small that its presence disturbs only the
second order term $S_2 (x)$ of the JWKB approximation, which does not
appear in the first order semiclassical solution~(\ref{escsol}), but
it does turn up in the condition~(\ref{esccondb}) that defines the
validity region $I$ of the semiclassical approximation. So it is safer
anyway to get rid of the term $\delta^2 W_0 (x)$ before applying the
JWKB method.

Finally we note that in the works
\cite{CsoGraSze97,FliCsoGraSze97,CsoGraSze98} the same physical
problem (collective excitations in Bose-condensed gases in spherically
symmetric harmonic traps) has been studied with basically different
semiclassical methods. The authors of these works could handle also
the $l \ne 0$ case. They applied the Langer modification \cite{Lan37}
$l(l+1)\to \bigl(l+\frac{1}{2} \bigr)^2$, and in their formula the
shift of the energy square is $1$ in dimensionless units, as compared
to the exact result.

As a next step, it would be useful to extend the present semiclassical
method also for the $l \ne 0$ case, and then apply the same techniques
for the solution of the Stringari equation in non-harmonic potentials,
which are analytically not treatable, but which have more and more
experimental relevance with the growth of the extent of the
condensate.

\acknowledgments

The author is thankful to Professor P\'eter Sz\'epfalusy for proposing
the theme of this work, for his ideas and his reading the manuscript,
and also indebted to Professor Andr\'e Voros for his pieces of advice, for the long and helpful discussions and for his kind hospitality.

This work has been supported by the Hungarian Academy of Sciences
under Grant No.\ AKP~98-20, by the Hungarian National Scientific
Research Foundation under Grant No.\ OTKA~T029552 and by the National
Committee for Technological Development under Grant No.\ 
OMFB~EU-98/155.

\bibliography{cikkstr_a_4}

\end{document}